\documentclass[twocolumn,prb,amsmath,amssymb,floatfix,superscriptaddress,showpacs]{revtex4}
\usepackage{graphicx}
\usepackage{dcolumn}
\usepackage{bm}
\usepackage{times}
\begin{document}

\title{Short-range incommensurate magnetic order near the superconducting phase boundary in Fe$_{1+\delta}$Te$_{1-x}$Se$_x$}
\author{Jinsheng Wen}
\affiliation{Condensed Matter Physics and Materials Science
Department, Brookhaven National Laboratory, Upton, NY 11973, USA}
\affiliation{Department of Materials Science and Engineering, Stony
Brook University, Stony Brook, NY 11794, USA}
\author{Guangyong Xu}
\affiliation{Condensed Matter Physics and Materials Science
Department, Brookhaven National Laboratory, Upton, NY 11973, USA}
\author{Zhijun Xu}
\affiliation{Condensed Matter Physics and Materials Science
Department, Brookhaven National Laboratory, Upton, NY 11973, USA}
\affiliation{Physics Department, the City College of New York, New
York, NY 10031, USA}
\author{Zhi Wei Lin}
\affiliation{Condensed Matter Physics and Materials Science
Department, Brookhaven National Laboratory, Upton, NY 11973, USA}
\author{Qiang Li}
\affiliation{Condensed Matter Physics and Materials Science
Department, Brookhaven National Laboratory, Upton, NY 11973, USA}
\author{W. Ratcliff}
\affiliation{NIST Center for Neutron Research, National Institute of
Standards and Technology, Gaithersburg, Maryland 20899, USA}
\author{Genda Gu}
\affiliation{Condensed Matter Physics and Materials Science
Department, Brookhaven National Laboratory, Upton, NY 11973, USA}
\author{J.~M.~Tranquada}
\affiliation{Condensed Matter Physics and Materials Science
Department, Brookhaven National Laboratory, Upton, NY 11973, USA}
\date{\today}

\begin{abstract}
We performed elastic neutron scattering and magnetization
measurements on Fe$_{1.07}$Te$_{0.75}$Se$_{0.25}$ and
FeTe$_{0.7}$Se$_{0.3}$. Short-range incommensurate magnetic order is
observed in both samples. In the former sample with higher Fe
content, a broad magnetic peak appears around  (0.46,0,0.5) at low
temperature, while in FeTe$_{0.7}$Se$_{0.3}$ the broad magnetic peak
is found to be closer to the antiferromagnetic (AFM) wave-vector
(0.5,0,0.5). The incommensurate peaks are only observed on one side
of the AFM wave-vector for both samples, which can be modeled in
terms of an imbalance of ferromagnetic/antiferromagnetic
correlations between nearest-neighbor spins. We also find that with
higher Se (and lower Fe) concentration, the magnetic order becomes
weaker while the superconducting temperature and volume increase.
\end{abstract}


\maketitle

Since the recent discovery of Fe-based superconductors with high
critical temperatures
($T_c$),~\cite{hosono_2,hosono4,chen-2008-453,ren-2008-25} extensive
research has been carried out to study the magnetic structures in
these materials,~\cite{cruz,qiu:257002,huang:257003} as magnetic
fluctuations are expected to play an important role in producing the
unconventional
superconductivity.~\cite{ma:033111,dong-2008-83,mazin:057003} It is
now well established that in LaFeAsO-(1:1:1:1)
(Refs.~\onlinecite{zhao,luetkens-2008,drew-2009}) and
BaFe$_2$As$_2$-(1:2:2)
(Refs.~\onlinecite{rotter-2008-47,chen-2009-85,fang-2009b,chu:014506})
type compounds, the long-range magnetic order is suppressed with
doping, while the superconductivity appears above a certain doping
value. While there are some rare cases where superconductivity
appears sharply after magnetic order
disappears,~\cite{luetkens-2008} in most systems short-range
magnetic order coexists with superconductivity over some range of
doping.~\cite{zhao,drew-2009,rotter-2008-47,chen-2009-85,fang-2009b,chu:014506}

In the more recently discovered system
Fe$_{1+\delta}$Te$_{1-x}$Se$_x$
(1:1),\cite{hsu-2008,yeh-2008,fang-2008-78} it is found that: i)
long-range magnetic order is present in non-superconducting
Fe$_{1+\delta}$Te,~\cite{fruchart-1975,li-2009-79,bao-2008} but only
short-range magnetic order survives in superconducting samples with
33\%~(Ref.~\onlinecite{bao-2008}) and 40\% Se
(Ref.~\onlinecite{fang-2008-78}); ii) the observed magnetic order
has a different propagation wave-vector from that of the other
Fe-based systems. To describe the ordering, we consider a tetragonal
unit cell containing two Fe atoms per plane, and specify wave
vectors in  reciprocal lattice units (r.l.u.) of $(a^*, b^*,
c^*)=(2\pi/a,2\pi/b,2\pi/c)$; the unit cell is rotated $45^\circ$ in
the $a$--$b$ plane from that used for 1:1:1:1 and 1:2:2
systems.~\cite{li-2009-79} In the latter systems, the
spin-density-wave (SDW) order is commensurate, with propagation
wave-vector (0.5,0.5,0.5), generally attributed to nesting of the
Fermi surface.~\cite{li-2009-79,zhao:140504,maier:020514,yin:047001}
In the 1:1 system, the SDW order propagates along (0.5,0,0.5), and
can be either commensurate, or incommensurate, depending on the Fe
content.~\cite{li-2009-79,bao-2008} Calculations using the local
spin density approximation for hypothetical stoichiometric FeTe
yield a commensurate magnetic ground state consistent with that seen
experimentally\cite{ma-2009,johannes-2009}; however, the
(0.5,0.5,0.5) SDW order is calculated to have the lowest energy for
FeSe.\cite{ma-2009}

To address the evolution of the magnetic correlations with Se
concentration, we have performed elastic neutron scattering and
magnetization measurements on high quality single crystals with
different Fe and Se contents. We show that there is short-range
incommensurate magnetic order in both
Fe$_{1.07}$Te$_{0.75}$Se$_{0.25}$ and FeTe$_{0.7}$Se$_{0.3}$ at low
temperature. Broad magnetic peaks appear at positions slightly
displaced from the antiferromagnetic (AFM) wave-vector $(0.5,0,0.5)$
in both samples when cooled below $\alt40$~K. The peak intensity
increases with further cooling and persists into the superconducting
phase. The magnetic peak intensity drops with more Se and less Fe
content, and with strengthening superconductivity.

Single crystals with nice (001) cleavage planes were grown by a
unidirectional solidification method with nominal compositions of
Fe$_{1.07}$Te$_{0.75}$Se$_{0.25}$ and FeTe$_{0.7}$Se$_{0.3}$ and
respective masses of 4.7 and 7.2~g. Neutron scattering experiments
were carried out on the triple-axis spectrometer BT-9 located at the
NIST Center for Neutron Research. The scattering plane $(H0L)$ is
defined by two vectors [100] and [001] in tetragonal notation. The
lattice constants for both samples are $a=b=3.80(8)$~\AA, and
$c=6.14(7)$~\AA.

\begin{figure}[ht]
\includegraphics[width=0.8\linewidth]{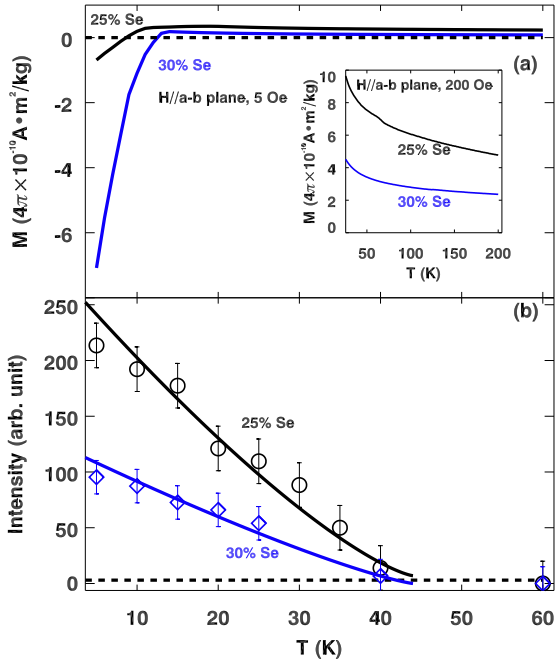}
\caption{(Color online) (a) ZFC magnetization, and (b) background
subtracted magnetic peak intensity measured along [100] (normalized
to the sample mass) as a function of temperature for
Fe$_{1.07}$Te$_{0.75}$Se$_{0.25}$, and FeTe$_{0.7}$Se$_{0.3}$. Error
bars indicate one standard deviation assuming Poisson
statistics. Lines through data are guides for the eyes.}
\label{fig:1}
\end{figure}

The bulk magnetization was characterized using a superconducting
quantum interference device (SQUID) magnetometer. In the
magnetization measurements, each sample was oriented so that the
(001) plane was parallel to the magnetic field. The
zero-field-cooling (ZFC) magnetization {\it vs.} temperature for
each sample is shown in Fig.~\ref{fig:1}(a), where one can see that
the 25\%\ Se sample only shows a trace of superconductivity, while
the 30\%\ Se sample clearly has a $T_c\sim13$~K.  We estimate that
the superconducting volume fraction for the latter sample is
$\sim1$\%. The inset of Fig.~1(a) shows that the paramagnetic
magnetization grows on cooling, and is greater in the sample with
less Se (and more Fe). The paramagnetic response does not follow
simple  Curie-Weiss behavior, so it is not possible to make a
meaningful estimate of effective magnetic moments.  For the 25\%\ Se
sample, there is a shoulder at $\sim60$~K which could be due to
2--3\%\ of Fe$_{1+\delta}$Te as a second phase, which has a magnetic
phase transition temperature of $\sim$65~K.~\cite{li-2009-79}

\begin{figure}[ht]
\includegraphics[width=0.9\linewidth]{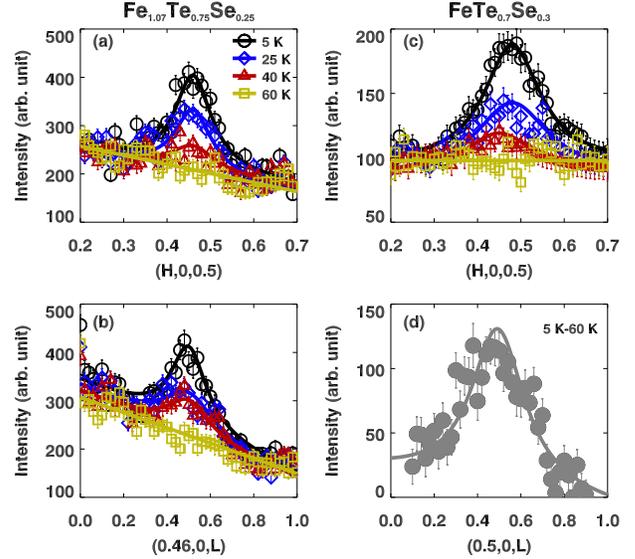}
\caption{(Color online) Short-range magnetic order in
Fe$_{1+\delta}$Te$_{1-x}$Se$_x$. The left and right columns show the
magnetic peak profiles for Fe$_{1.07}$Te$_{0.75}$Se$_{0.25}$ and
FeTe$_{0.7}$Se$_{0.3}$, respectively. Top and bottom rows are scans
along [100] and [001] respectively. (a), (b), and (c) are data taken
at various temperatures. For the 30\% Se
sample, there is a temperature-independent spurious peak in
the [001] scans, so in (d) we only plot 5~K data with the 60-K scan subtracted.
All data are taken with 1
minute counting time and then normalized to the sample mass. Error
bars represent the square root of the total counts. The lines are
fits to the data using Lorentzian functions.} \label{fig:2}
\end{figure}

In our elastic neutron scattering measurements,  each sample was
aligned on the (200) and (001) nuclear Bragg peaks with an accuracy
and reproducibility in longitudinal wave vector of better than
0.005~r.l.u..  For the magnetic peaks, linear scans were performed
along [100] and [001] directions at various temperatures. The
temperature dependence of the peak intensity is summarized in
Fig.~\ref{fig:1}(b), and representative scans are shown in
Fig.~\ref{fig:2}. No net peak intensity is observed at 60~K, but a
weak magnetic peak appears at slightly lower temperature, growing in
intensity with further cooling. For
Fe$_{1.07}$Te$_{0.75}$Se$_{0.25}$, the magnetic structure is clearly
incommensurate, and the peak position is determined to be
($0.5-\epsilon,0,0.5$), with $\epsilon=0.04$. From
Fig.~\ref{fig:2}(a), we did not observe a peak at
($0.5+\epsilon,0,0.5$). For FeTe$_{0.7}$Se$_{0.3}$, the magnetic
peak center is at (0.48,0,0.5), although this differs from the
commensurate position by less than the peak width. Our observations
are qualitatively consistent with the previous result\cite{bao-2008}
for Fe$_{1.08}$Te$_{0.67}$Se$_{0.33}$, where the magnetic peak is at
(0.438,0,0.5); it appears that both the Fe and Se concentrations
impact the ordering wave-vector. We have also searched for SDW order
around (0.5,0.5,0.5) in the $(HHL)$ zone, but no evidence of
magnetic peaks was found.

At 5~K, the peak width for Fe$_{1.07}$Te$_{0.75}$Se$_{0.25}$ [100]
scan is 0.10~r.l.u., which corresponds to a correlation length of
6.1(1)~\AA. The width along [001] is 0.20~r.l.u.,  giving a
correlation length of 4.9(1)~\AA. As can be seen from
Fig.~\ref{fig:2}, the peaks for FeTe$_{0.7}$Se$_{0.3}$ along [100]
and [001] are broader than their counterparts for
Fe$_{1.07}$Te$_{0.75}$Se$_{0.25}$, and the correlation lengths are
determined to be 3.8(1)~\AA{} along [100] and 3.3(1)~\AA{} along
[001]. Also, from Fig.~\ref{fig:1}(b), one can see that the magnetic
peak intensity for Fe$_{1.07}$Te$_{0.75}$Se$_{0.25}$ is always
higher than the other one. Although the SDW order is short-ranged in
both compounds, and starts at around the same temperature,
$\sim40$~K, the order is apparently stronger in the 25\% Se sample.

\begin{figure}[ht]
\includegraphics[width=0.8\linewidth]{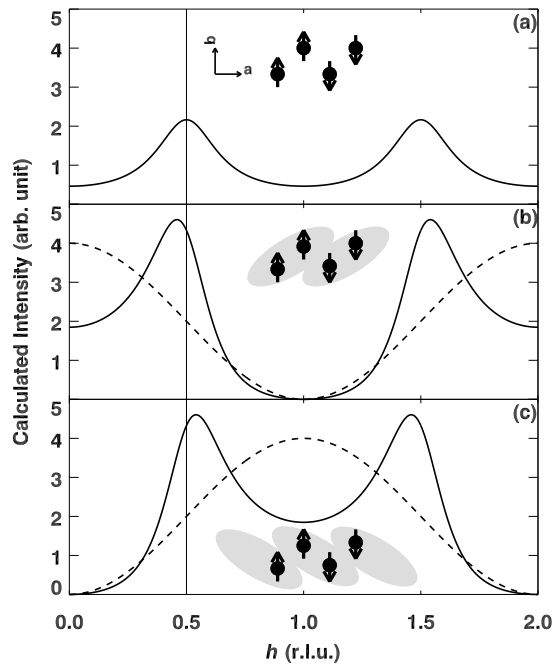}
\caption{(a) Inset shows the commensurate magnetic unit cell within
a single layer of Fe$_{1+\delta}$Te, with spin arrangements in
$a$-$b$ plane; solid line shows the calculated scattered intensity
assuming uniform exponential decay of spin correlations. (b) Dashed
line shows the magnetic structure factor $|F|^2$ and solid line
shows calculated intensity for exponential decay of correlations
between ferromagnetic spin pairs (inset). (c) Same as (b) but for
exponential decay of correlations between antiferromagnetic spin
pairs.} \label{fig:3}
\end{figure}

The magnetic structure of the parent compound Fe$_{1+\delta}$Te can
be described by the schematic diagram in the inset of
Fig.~\ref{fig:3}(a), which is adopted from
Refs.~\onlinecite{li-2009-79,bao-2008}. Here the magnetic structure
consists of two spin sublattices. The spins in both sublattices are
found to be aligned along $b$-axis. Within each sublattice, the
spins have an antiferromagnetic alignment along $a$ and $c$-axes,
and ferromagnetic along the $b$-axis. The spins have a small
out-of-plane component, but here, for simplicity, we are only
considering the components in the $a$--$b$ plane. With low excess
Fe,~\cite{li-2009-79} this configuration gives rise to magnetic
Bragg peaks at the commensurate AFM wave-vector (0.5,0,0.5). The
extra Fe is considered to reside in the interstitial sites of the
Te/Se atoms.~\cite{bao-2008} With more excess Fe, the ordering
wave-vector becomes incommensurate, which can be explained by a
modulation of the ordered moment size and orientation, propagating
along the $a$-axis.~\cite{bao-2008} The connection between excess Fe and the transition
from commensurate to incommensurate order has been modeled theoretically.~\cite{fang-2009}

With Se doping, the magnetic order is depressed and becomes short
ranged. It is intriguing that magnetic order can survive without a
lowering of the lattice symmetry from tetragonal, although perhaps
there are local symmetry reductions on the scale of the magnetic
correlation length. The incommensurability is also interesting. A
uniform sinusoidal modulation of the spin directions or magnitudes
will give incommensurate peaks at $(0.5\pm\epsilon,0,0.5)$, whereas
we see a peak only on the $-\epsilon$ side. One can model this with
phase shifted modulations on the two sublattices, but the modulation
length required to describe the incommensurability is much greater
than the correlation length.

We have found that a simple description of the incommensurability
can be obtained when the decay of correlations between ferromagnetic
nearest-neighbor spins  is different from that of antiferromagnetic
spin neighbors.  We will consider correlations only along the
modulation direction within an $a$--$b$ plane, and assume that they
are independent of correlations in the orthogonal directions.  Let
us break the spin system into perfectly correlated nearest-neighbor
pairs, with exponential decay of the spin correlations from one pair
to the next along the $a$-axis.  The neutron scattering intensity
can then be expressed as~\cite{guinier-1994}
\begin{equation}\label{eq1}
    I \propto |F|^2\frac{1-p^2}{1+p^2-2p\cos(2\pi h)},
\end{equation}
where $F$ is the structure factor for the selected pair of spins,
$h$ is the wave-vector component along the $a$-axis, and
\begin{equation}\label{eq2}
    p = -e^{-a/\xi};
\end{equation}
 $p$ is the correlation function between neighboring pairs, where the negative
sign suggests that the inter-pair corrlation is antiferromagnetic;
and $\xi$ is the
correlation length. (In all cases discussed below, we set $\xi=a$.)

Let us first consider the case of ferromagnetic spin pairs with
exponentially decaying correlations between pairs, as illustrated in
Fig.~\ref{fig:3}(b). The structure factor for this case corresponds
to
\begin{equation}\label{eq3}
    |F|^2=4\cos^2({\textstyle\frac12}\pi h),
\end{equation}
as indicated by the dashed line in Fig.~\ref{fig:3}(b). Plugging
this into Eq.~(\ref{eq1}) gives the solid line shown in
Fig.~\ref{fig:3}(b).  Note that the calculated peaks are
incommensurate, with the peak near $h=0.5$ shifted to lower $h$.
Alternatively, we can start with an antiferromagnetic spin pair, in
which case
\begin{equation}\label{eq4}
    |F|^2=4\sin^2({\textstyle\frac12}\pi h).
\end{equation}
This yields the result shown in Fig.~\ref{fig:3}(c), with the peaks
shifted in the opposite direction.  If the decay of correlations is
identical for ferromagnetic and antiferromagnetic nearest neighbors,
then we can average over these two cases, obtaining $|F|^2=2$; the
resulting commensurate peaks are shown in Fig.~\ref{fig:3}(a).

Our experimental results look similar to Fig.~\ref{fig:3}(b). This
suggests that the ferromagnetic correlations are stronger than the
antiferromagnetic ones.  For the model illustrated in
Fig.~\ref{fig:3}(b), the incommensurability grows as the correlation
length gets shorter. The trend in our two samples does not follow
this relationship; however, one could describe a more general
relationship between the ferromagnetic and antiferromagnetic
correlations by taking a weighted average of Eqs.~(\ref{eq3}) and
(\ref{eq4}).

In summary, we have observed short-range magnetic order in
Fe$_{1.07}$Te$_{0.75}$Se$_{0.25}$ and FeTe$_{0.7}$Se$_{0.3}$. In
both samples, the magnetic order is incommensurate and only observed
on one side of the commensurate wave-vector (0.5,0,0.5), which is
likely a result of the imbalance of ferromagnetic/antiferromagnetic
correlations between neighboring spins. The parent compound
Fe$_{1+\delta}$Te is not superconducting,~\cite{bao-2008,li-2009-79}
and the optimally doped sample with 50\% Se has no static magnetic
order~\cite{qiu-2009,wen-2009}. Our samples have Se content lying in
the middle, where we see that with larger Se doping, the SDW order
becomes weaker, while the superconductivity is enhanced. This could
imply the coexistence and competition between SDW order and
superconductivity in this system, similar to other
Fe-based~\cite{dong-2008-83,zhao,drew-2009,rotter-2008-47,chen-2009-85}
and cuprate superconductors~\cite{prl67202,moodenbaugh,lakefield}.
Interestingly, in the Fe$_{1+\delta}$Te$_{1-x}$Se$_x$ system, the
SDW order and superconductivity can be tuned not only by doping Se,
but also by adjusting the Fe
content.~\cite{zhang:012506,fang-2008-78,mcqueen:014522} It has been
reported that the excess Fe acts as a magnetic electron
donor,~\cite{zhang:012506} suppresses the superconductivity, and
induces a weakly localized electronic state.~\cite{liu} Our results
are completely consistent with these results---with less Fe and more
Se, the SDW order is weaker; with more excess Fe and less Se,
superconductivity is weaker, but to really distinguish the role of
Fe and Se, samples only varying one element are certainly required
for future work. We also note that recent studies of superconducting
FeTe$_{0.6}$Se$_{0.4}$ (Ref.~\onlinecite{qiu-2009}) and
FeTe$_{0.5}$Se$_{0.5}$ (Ref.~\onlinecite{mook-2009}) show evidence
of a spin gap and resonance peak at the wave vector $(0.5,0.5,L)$.
It should be interesting to study how the magnetic correlations
evolve with Se concentration between 30\%\ and 40\%.

The work at Brookhaven National Laboratory was supported by the
Office of Science, U.S. Department of Energy, under Contract
No.\,DE-AC02-98CH10886.


\begin{thebibliography}{39}
\expandafter\ifx\csname natexlab\endcsname\relax\def\natexlab#1{#1}\fi
\expandafter\ifx\csname bibnamefont\endcsname\relax
  \def\bibnamefont#1{#1}\fi
\expandafter\ifx\csname bibfnamefont\endcsname\relax
  \def\bibfnamefont#1{#1}\fi
\expandafter\ifx\csname citenamefont\endcsname\relax
  \def\citenamefont#1{#1}\fi
\expandafter\ifx\csname url\endcsname\relax
  \def\url#1{\texttt{#1}}\fi
\expandafter\ifx\csname urlprefix\endcsname\relax\def\urlprefix{URL }\fi
\providecommand{\bibinfo}[2]{#2}
\providecommand{\eprint}[2][]{\url{#2}}

\bibitem[{\citenamefont{Kamihara et~al.}(2008)\citenamefont{Kamihara, Watanabe,
  Hirano, and Hosono}}]{hosono_2}
\bibinfo{author}{\bibfnamefont{Y.}~\bibnamefont{Kamihara}},
  \bibinfo{author}{\bibfnamefont{T.}~\bibnamefont{Watanabe}},
  \bibinfo{author}{\bibfnamefont{M.}~\bibnamefont{Hirano}}, \bibnamefont{and}
  \bibinfo{author}{\bibfnamefont{H.}~\bibnamefont{Hosono}},
  \bibinfo{journal}{J. Am. Chem. Soc.} \textbf{\bibinfo{volume}{130}},
  \bibinfo{pages}{3296} (\bibinfo{year}{2008}).

\bibitem[{\citenamefont{Takahashi et~al.}(2008)\citenamefont{Takahashi, Igawa,
  Arii, Kamihara, Hirano, and Hosono}}]{hosono4}
\bibinfo{author}{\bibfnamefont{H.}~\bibnamefont{Takahashi}},
  \bibinfo{author}{\bibfnamefont{K.}~\bibnamefont{Igawa}},
  \bibinfo{author}{\bibfnamefont{K.}~\bibnamefont{Arii}},
  \bibinfo{author}{\bibfnamefont{Y.}~\bibnamefont{Kamihara}},
  \bibinfo{author}{\bibfnamefont{M.}~\bibnamefont{Hirano}}, \bibnamefont{and}
  \bibinfo{author}{\bibfnamefont{H.}~\bibnamefont{Hosono}},
  \bibinfo{journal}{Nature} \textbf{\bibinfo{volume}{453}},
  \bibinfo{pages}{376} (\bibinfo{year}{2008}).

\bibitem[{\citenamefont{Chen et~al.}(2008)\citenamefont{Chen, Wu, Wu, Liu,
  Chen, and Fang}}]{chen-2008-453}
\bibinfo{author}{\bibfnamefont{X.~H.} \bibnamefont{Chen}},
  \bibinfo{author}{\bibfnamefont{T.}~\bibnamefont{Wu}},
  \bibinfo{author}{\bibfnamefont{G.}~\bibnamefont{Wu}},
  \bibinfo{author}{\bibfnamefont{R.~H.} \bibnamefont{Liu}},
  \bibinfo{author}{\bibfnamefont{H.}~\bibnamefont{Chen}}, \bibnamefont{and}
  \bibinfo{author}{\bibfnamefont{D.~F.} \bibnamefont{Fang}},
  \bibinfo{journal}{Nature} \textbf{\bibinfo{volume}{453}},
  \bibinfo{pages}{761} (\bibinfo{year}{2008}).

\bibitem[{\citenamefont{Ren et~al.}(2008)\citenamefont{Ren, Lu, Yang, Yi, Shen,
  Li, Che, Dong, Sun, Zhou et~al.}}]{ren-2008-25}
\bibinfo{author}{\bibfnamefont{Z.-A.} \bibnamefont{Ren}},
  \bibinfo{author}{\bibfnamefont{W.}~\bibnamefont{Lu}},
  \bibinfo{author}{\bibfnamefont{J.}~\bibnamefont{Yang}},
  \bibinfo{author}{\bibfnamefont{W.}~\bibnamefont{Yi}},
  \bibinfo{author}{\bibfnamefont{X.-L.} \bibnamefont{Shen}},
  \bibinfo{author}{\bibfnamefont{Z.-C.} \bibnamefont{Li}},
  \bibinfo{author}{\bibfnamefont{G.-C.} \bibnamefont{Che}},
  \bibinfo{author}{\bibfnamefont{X.-L.} \bibnamefont{Dong}},
  \bibinfo{author}{\bibfnamefont{L.-L.} \bibnamefont{Sun}},
  \bibinfo{author}{\bibfnamefont{F.}~\bibnamefont{Zhou}}, \bibnamefont{et~al.},
  \bibinfo{journal}{Chin. Phys. Lett.} \textbf{\bibinfo{volume}{25}},
  \bibinfo{pages}{2215} (\bibinfo{year}{2008}).

\bibitem[{\citenamefont{de~la Cruz et~al.}(2008)\citenamefont{de~la Cruz,
  Huang, Lynn, Li, Ratcliff, Zarestzky, Mook, Chen, Luo, Wang et~al.}}]{cruz}
\bibinfo{author}{\bibfnamefont{C.}~\bibnamefont{de~la Cruz}},
  \bibinfo{author}{\bibfnamefont{Q.}~\bibnamefont{Huang}},
  \bibinfo{author}{\bibfnamefont{J.~W.} \bibnamefont{Lynn}},
  \bibinfo{author}{\bibfnamefont{J.}~\bibnamefont{Li}},
  \bibinfo{author}{\bibfnamefont{W.}~\bibnamefont{Ratcliff}},
  \bibinfo{author}{\bibfnamefont{J.~L.} \bibnamefont{Zarestzky}},
  \bibinfo{author}{\bibfnamefont{H.~A.} \bibnamefont{Mook}},
  \bibinfo{author}{\bibfnamefont{G.~F.} \bibnamefont{Chen}},
  \bibinfo{author}{\bibfnamefont{J.~L.} \bibnamefont{Luo}},
  \bibinfo{author}{\bibfnamefont{N.~L.} \bibnamefont{Wang}},
  \bibnamefont{et~al.}, \bibinfo{journal}{Nature}
  \textbf{\bibinfo{volume}{453}}, \bibinfo{pages}{899} (\bibinfo{year}{2008}).

\bibitem[{\citenamefont{Qiu et~al.}(2008)\citenamefont{Qiu, Bao, Huang,
  Yildirim, Simmons, Green, Lynn, Gasparovic, Li, Wu et~al.}}]{qiu:257002}
\bibinfo{author}{\bibfnamefont{Y.}~\bibnamefont{Qiu}},
  \bibinfo{author}{\bibfnamefont{W.}~\bibnamefont{Bao}},
  \bibinfo{author}{\bibfnamefont{Q.}~\bibnamefont{Huang}},
  \bibinfo{author}{\bibfnamefont{T.}~\bibnamefont{Yildirim}},
  \bibinfo{author}{\bibfnamefont{J.~M.} \bibnamefont{Simmons}},
  \bibinfo{author}{\bibfnamefont{M.~A.} \bibnamefont{Green}},
  \bibinfo{author}{\bibfnamefont{J.~W.} \bibnamefont{Lynn}},
  \bibinfo{author}{\bibfnamefont{Y.~C.} \bibnamefont{Gasparovic}},
  \bibinfo{author}{\bibfnamefont{J.}~\bibnamefont{Li}},
  \bibinfo{author}{\bibfnamefont{T.}~\bibnamefont{Wu}}, \bibnamefont{et~al.},
  \bibinfo{journal}{Phys. Rev. Lett.} \textbf{\bibinfo{volume}{101}},
  \bibinfo{pages}{257002} (\bibinfo{year}{2008}).

\bibitem[{\citenamefont{Huang et~al.}(2008)\citenamefont{Huang, Qiu, Bao,
  Green, Lynn, Gasparovic, Wu, Wu, and Chen}}]{huang:257003}
\bibinfo{author}{\bibfnamefont{Q.}~\bibnamefont{Huang}},
  \bibinfo{author}{\bibfnamefont{Y.}~\bibnamefont{Qiu}},
  \bibinfo{author}{\bibfnamefont{W.}~\bibnamefont{Bao}},
  \bibinfo{author}{\bibfnamefont{M.~A.} \bibnamefont{Green}},
  \bibinfo{author}{\bibfnamefont{J.~W.} \bibnamefont{Lynn}},
  \bibinfo{author}{\bibfnamefont{Y.~C.} \bibnamefont{Gasparovic}},
  \bibinfo{author}{\bibfnamefont{T.}~\bibnamefont{Wu}},
  \bibinfo{author}{\bibfnamefont{G.}~\bibnamefont{Wu}}, \bibnamefont{and}
  \bibinfo{author}{\bibfnamefont{X.~H.} \bibnamefont{Chen}},
  \bibinfo{journal}{Phys. Rev. Lett.} \textbf{\bibinfo{volume}{101}},
  \bibinfo{pages}{257003} (\bibinfo{year}{2008}).

\bibitem[{\citenamefont{Ma and Lu}(2008)}]{ma:033111}
\bibinfo{author}{\bibfnamefont{F.}~\bibnamefont{Ma}} \bibnamefont{and}
  \bibinfo{author}{\bibfnamefont{Z.-Y.} \bibnamefont{Lu}},
  \bibinfo{journal}{Phys. Rev. B} \textbf{\bibinfo{volume}{78}},
  \bibinfo{pages}{033111} (\bibinfo{year}{2008}).

\bibitem[{\citenamefont{Dong et~al.}(2008)\citenamefont{Dong, Zhang, Xu, Li,
  Li, Hu, Wu, Chen, Dai, Luo et~al.}}]{dong-2008-83}
\bibinfo{author}{\bibfnamefont{J.}~\bibnamefont{Dong}},
  \bibinfo{author}{\bibfnamefont{H.~J.} \bibnamefont{Zhang}},
  \bibinfo{author}{\bibfnamefont{G.}~\bibnamefont{Xu}},
  \bibinfo{author}{\bibfnamefont{Z.}~\bibnamefont{Li}},
  \bibinfo{author}{\bibfnamefont{G.}~\bibnamefont{Li}},
  \bibinfo{author}{\bibfnamefont{W.~Z.} \bibnamefont{Hu}},
  \bibinfo{author}{\bibfnamefont{D.}~\bibnamefont{Wu}},
  \bibinfo{author}{\bibfnamefont{G.~F.} \bibnamefont{Chen}},
  \bibinfo{author}{\bibfnamefont{X.}~\bibnamefont{Dai}},
  \bibinfo{author}{\bibfnamefont{J.~L.} \bibnamefont{Luo}},
  \bibnamefont{et~al.}, \bibinfo{journal}{Europhys. Lett.}
  \textbf{\bibinfo{volume}{83}}, \bibinfo{pages}{27006} (\bibinfo{year}{2008}).

\bibitem[{\citenamefont{Mazin et~al.}(2008)\citenamefont{Mazin, Singh,
  Johannes, and Du}}]{mazin:057003}
\bibinfo{author}{\bibfnamefont{I.~I.} \bibnamefont{Mazin}},
  \bibinfo{author}{\bibfnamefont{D.~J.} \bibnamefont{Singh}},
  \bibinfo{author}{\bibfnamefont{M.~D.} \bibnamefont{Johannes}},
  \bibnamefont{and} \bibinfo{author}{\bibfnamefont{M.~H.} \bibnamefont{Du}},
  \bibinfo{journal}{Phys. Rev. Lett.} \textbf{\bibinfo{volume}{101}},
  \bibinfo{pages}{057003} (\bibinfo{year}{2008}).

\bibitem[{\citenamefont{Zhao et~al.}(2008{\natexlab{a}})\citenamefont{Zhao,
  Huang, Cruz, Li, Lynn, Chen, Green, Chen, Li, Li et~al.}}]{zhao}
\bibinfo{author}{\bibfnamefont{J.}~\bibnamefont{Zhao}},
  \bibinfo{author}{\bibfnamefont{Q.}~\bibnamefont{Huang}},
  \bibinfo{author}{\bibfnamefont{C.~D.~L.} \bibnamefont{Cruz}},
  \bibinfo{author}{\bibfnamefont{S.}~\bibnamefont{Li}},
  \bibinfo{author}{\bibfnamefont{J.~W.} \bibnamefont{Lynn}},
  \bibinfo{author}{\bibfnamefont{Y.}~\bibnamefont{Chen}},
  \bibinfo{author}{\bibfnamefont{M.~A.} \bibnamefont{Green}},
  \bibinfo{author}{\bibfnamefont{G.~F.} \bibnamefont{Chen}},
  \bibinfo{author}{\bibfnamefont{G.}~\bibnamefont{Li}},
  \bibinfo{author}{\bibfnamefont{Z.}~\bibnamefont{Li}}, \bibnamefont{et~al.},
  \bibinfo{journal}{Nature Mater.} \textbf{\bibinfo{volume}{7}},
  \bibinfo{pages}{953} (\bibinfo{year}{2008}{\natexlab{a}}).

\bibitem[{\citenamefont{Luetkens et~al.}(2009)\citenamefont{Luetkens, Klauss,
  Kraken, Litterst, Dellmann, Klingeler, Hess, Khasanov, Amato, Baines
  et~al.}}]{luetkens-2008}
\bibinfo{author}{\bibfnamefont{H.}~\bibnamefont{Luetkens}},
  \bibinfo{author}{\bibfnamefont{H.~H.} \bibnamefont{Klauss}},
  \bibinfo{author}{\bibfnamefont{M.}~\bibnamefont{Kraken}},
  \bibinfo{author}{\bibfnamefont{F.~J.} \bibnamefont{Litterst}},
  \bibinfo{author}{\bibfnamefont{T.}~\bibnamefont{Dellmann}},
  \bibinfo{author}{\bibfnamefont{R.}~\bibnamefont{Klingeler}},
  \bibinfo{author}{\bibfnamefont{C.}~\bibnamefont{Hess}},
  \bibinfo{author}{\bibfnamefont{R.}~\bibnamefont{Khasanov}},
  \bibinfo{author}{\bibfnamefont{A.}~\bibnamefont{Amato}},
  \bibinfo{author}{\bibfnamefont{C.}~\bibnamefont{Baines}},
  \bibnamefont{et~al.}, \bibinfo{journal}{Nature Mater.}
  \textbf{\bibinfo{volume}{8}}, \bibinfo{pages}{305} (\bibinfo{year}{2009}).

\bibitem[{\citenamefont{Drew et~al.}(2009)\citenamefont{Drew, Niedermayer,
  Baker, Pratt, Blundell, Lancaster, Liu, Wu, Chen, Watanabe
  et~al.}}]{drew-2009}
\bibinfo{author}{\bibfnamefont{A.~J.} \bibnamefont{Drew}},
  \bibinfo{author}{\bibfnamefont{C.}~\bibnamefont{Niedermayer}},
  \bibinfo{author}{\bibfnamefont{P.~J.} \bibnamefont{Baker}},
  \bibinfo{author}{\bibfnamefont{F.~L.} \bibnamefont{Pratt}},
  \bibinfo{author}{\bibfnamefont{S.~J.} \bibnamefont{Blundell}},
  \bibinfo{author}{\bibfnamefont{T.}~\bibnamefont{Lancaster}},
  \bibinfo{author}{\bibfnamefont{R.~H.} \bibnamefont{Liu}},
  \bibinfo{author}{\bibfnamefont{G.}~\bibnamefont{Wu}},
  \bibinfo{author}{\bibfnamefont{X.~H.} \bibnamefont{Chen}},
  \bibinfo{author}{\bibfnamefont{I.}~\bibnamefont{Watanabe}},
  \bibnamefont{et~al.}, \bibinfo{journal}{Nature Mater.}
  \textbf{\bibinfo{volume}{8}}, \bibinfo{pages}{310} (\bibinfo{year}{2009}).

\bibitem[{\citenamefont{Rotter et~al.}(2008)\citenamefont{Rotter, Pangerl,
  Tegel, and Johrendt}}]{rotter-2008-47}
\bibinfo{author}{\bibfnamefont{M.}~\bibnamefont{Rotter}},
  \bibinfo{author}{\bibfnamefont{M.}~\bibnamefont{Pangerl}},
  \bibinfo{author}{\bibfnamefont{M.}~\bibnamefont{Tegel}}, \bibnamefont{and}
  \bibinfo{author}{\bibfnamefont{D.}~\bibnamefont{Johrendt}},
  \bibinfo{journal}{Angew. Chem. Int.. Ed.} \textbf{\bibinfo{volume}{47}},
  \bibinfo{pages}{7949} (\bibinfo{year}{2008}).

\bibitem[{\citenamefont{Chen et~al.}(2009)\citenamefont{Chen, Ren, Qiu, bao,
  Liu, Wu, Wu, Xie, Wang, Huang et~al.}}]{chen-2009-85}
\bibinfo{author}{\bibfnamefont{H.}~\bibnamefont{Chen}},
  \bibinfo{author}{\bibfnamefont{Y.}~\bibnamefont{Ren}},
  \bibinfo{author}{\bibfnamefont{Y.}~\bibnamefont{Qiu}},
  \bibinfo{author}{\bibfnamefont{W.}~\bibnamefont{bao}},
  \bibinfo{author}{\bibfnamefont{R.~H.} \bibnamefont{Liu}},
  \bibinfo{author}{\bibfnamefont{G.}~\bibnamefont{Wu}},
  \bibinfo{author}{\bibfnamefont{T.}~\bibnamefont{Wu}},
  \bibinfo{author}{\bibfnamefont{Y.~L.} \bibnamefont{Xie}},
  \bibinfo{author}{\bibfnamefont{X.~F.} \bibnamefont{Wang}},
  \bibinfo{author}{\bibfnamefont{Q.}~\bibnamefont{Huang}},
  \bibnamefont{et~al.}, \bibinfo{journal}{Europhys. Lett.}
  \textbf{\bibinfo{volume}{85}}, \bibinfo{pages}{17006} (\bibinfo{year}{2009}).

\bibitem[{\citenamefont{Fang et~al.}(2009{\natexlab{a}})\citenamefont{Fang,
  Luo, Cheng, Wang, Jia, Mu, Shen, Mazin, Shan, Ren et~al.}}]{fang-2009b}
\bibinfo{author}{\bibfnamefont{L.}~\bibnamefont{Fang}},
  \bibinfo{author}{\bibfnamefont{H.}~\bibnamefont{Luo}},
  \bibinfo{author}{\bibfnamefont{P.}~\bibnamefont{Cheng}},
  \bibinfo{author}{\bibfnamefont{Z.}~\bibnamefont{Wang}},
  \bibinfo{author}{\bibfnamefont{Y.}~\bibnamefont{Jia}},
  \bibinfo{author}{\bibfnamefont{G.}~\bibnamefont{Mu}},
  \bibinfo{author}{\bibfnamefont{B.}~\bibnamefont{Shen}},
  \bibinfo{author}{\bibfnamefont{I.~I.} \bibnamefont{Mazin}},
  \bibinfo{author}{\bibfnamefont{L.}~\bibnamefont{Shan}},
  \bibinfo{author}{\bibfnamefont{C.}~\bibnamefont{Ren}}, \bibnamefont{et~al.},
  \eprint{arXiv:0903.2418}.

\bibitem[{\citenamefont{Chu et~al.}(2009)\citenamefont{Chu, Analytis,
  Kucharczyk, and Fisher}}]{chu:014506}
\bibinfo{author}{\bibfnamefont{J.-H.} \bibnamefont{Chu}},
  \bibinfo{author}{\bibfnamefont{J.~G.} \bibnamefont{Analytis}},
  \bibinfo{author}{\bibfnamefont{C.}~\bibnamefont{Kucharczyk}},
  \bibnamefont{and} \bibinfo{author}{\bibfnamefont{I.~R.}
  \bibnamefont{Fisher}}, \bibinfo{journal}{Phys. Rev. B}
  \textbf{\bibinfo{volume}{79}}, \bibinfo{pages}{014506}
  (\bibinfo{year}{2009}).

\bibitem[{\citenamefont{Hsu et~al.}(2008)\citenamefont{Hsu, Luo, Yeh, Chen,
  Huang, Wu, Lee, Huang, Chu, Yan et~al.}}]{hsu-2008}
\bibinfo{author}{\bibfnamefont{F.-C.} \bibnamefont{Hsu}},
  \bibinfo{author}{\bibfnamefont{J.-Y.} \bibnamefont{Luo}},
  \bibinfo{author}{\bibfnamefont{K.-W.} \bibnamefont{Yeh}},
  \bibinfo{author}{\bibfnamefont{T.-K.} \bibnamefont{Chen}},
  \bibinfo{author}{\bibfnamefont{T.-W.} \bibnamefont{Huang}},
  \bibinfo{author}{\bibfnamefont{P.~M.} \bibnamefont{Wu}},
  \bibinfo{author}{\bibfnamefont{Y.-C.} \bibnamefont{Lee}},
  \bibinfo{author}{\bibfnamefont{Y.-L.} \bibnamefont{Huang}},
  \bibinfo{author}{\bibfnamefont{Y.-Y.} \bibnamefont{Chu}},
  \bibinfo{author}{\bibfnamefont{D.-C.} \bibnamefont{Yan}},
  \bibnamefont{et~al.}, \bibinfo{journal}{Proc. Natl. Acad. Sci. U.S.A.}
  \textbf{\bibinfo{volume}{105}}, \bibinfo{pages}{14262}
  (\bibinfo{year}{2008}).

\bibitem[{\citenamefont{Yeh et~al.}(2008)\citenamefont{Yeh, Huang, Huang, Chen,
  Hsu, Wu, Lee, Chu, Chen, Luo et~al.}}]{yeh-2008}
\bibinfo{author}{\bibfnamefont{K.-W.} \bibnamefont{Yeh}},
  \bibinfo{author}{\bibfnamefont{T.-W.} \bibnamefont{Huang}},
  \bibinfo{author}{\bibfnamefont{Y.-L.} \bibnamefont{Huang}},
  \bibinfo{author}{\bibfnamefont{T.-K.} \bibnamefont{Chen}},
  \bibinfo{author}{\bibfnamefont{F.-C.} \bibnamefont{Hsu}},
  \bibinfo{author}{\bibfnamefont{P.~M.} \bibnamefont{Wu}},
  \bibinfo{author}{\bibfnamefont{Y.-C.} \bibnamefont{Lee}},
  \bibinfo{author}{\bibfnamefont{Y.-Y.} \bibnamefont{Chu}},
  \bibinfo{author}{\bibfnamefont{C.-L.} \bibnamefont{Chen}},
  \bibinfo{author}{\bibfnamefont{J.-Y.} \bibnamefont{Luo}},
  \bibnamefont{et~al.}, \bibinfo{journal}{Europhys. Lett.}
  \textbf{\bibinfo{volume}{84}}, \bibinfo{pages}{37002} (\bibinfo{year}{2008}).

\bibitem[{\citenamefont{Fang et~al.}(2008)\citenamefont{Fang, Pham, Qian, Liu,
  Vehstedt, Liu, Spinu, and Mao}}]{fang-2008-78}
\bibinfo{author}{\bibfnamefont{M.~H.} \bibnamefont{Fang}},
  \bibinfo{author}{\bibfnamefont{H.~M.} \bibnamefont{Pham}},
  \bibinfo{author}{\bibfnamefont{B.}~\bibnamefont{Qian}},
  \bibinfo{author}{\bibfnamefont{T.~J.} \bibnamefont{Liu}},
  \bibinfo{author}{\bibfnamefont{E.~K.} \bibnamefont{Vehstedt}},
  \bibinfo{author}{\bibfnamefont{Y.}~\bibnamefont{Liu}},
  \bibinfo{author}{\bibfnamefont{L.}~\bibnamefont{Spinu}}, \bibnamefont{and}
  \bibinfo{author}{\bibfnamefont{Z.~Q.} \bibnamefont{Mao}},
  \bibinfo{journal}{Phys. Rev. B} \textbf{\bibinfo{volume}{78}},
  \bibinfo{pages}{224503} (\bibinfo{year}{2008}).

\bibitem[{\citenamefont{Fruchart et~al.}(1975)\citenamefont{Fruchart, Convert,
  Wolfers, Madar, Senateur, and Fruchart}}]{fruchart-1975}
\bibinfo{author}{\bibfnamefont{D.}~\bibnamefont{Fruchart}},
  \bibinfo{author}{\bibfnamefont{P.}~\bibnamefont{Convert}},
  \bibinfo{author}{\bibfnamefont{P.}~\bibnamefont{Wolfers}},
  \bibinfo{author}{\bibfnamefont{R.}~\bibnamefont{Madar}},
  \bibinfo{author}{\bibfnamefont{J.~P.} \bibnamefont{Senateur}},
  \bibnamefont{and} \bibinfo{author}{\bibfnamefont{R.}~\bibnamefont{Fruchart}},
  \bibinfo{journal}{Mater. Res. Bull.} \textbf{\bibinfo{volume}{10}},
  \bibinfo{pages}{169 } (\bibinfo{year}{1975}).

\bibitem[{\citenamefont{Li et~al.}(2009)\citenamefont{Li, {de la Cruz}, Huang,
  Chen, Lynn, Hu, Huang, Hsu, Yeh, kuen Wu et~al.}}]{li-2009-79}
\bibinfo{author}{\bibfnamefont{S.}~\bibnamefont{Li}},
  \bibinfo{author}{\bibfnamefont{C.}~\bibnamefont{{de la Cruz}}},
  \bibinfo{author}{\bibfnamefont{Q.}~\bibnamefont{Huang}},
  \bibinfo{author}{\bibfnamefont{Y.}~\bibnamefont{Chen}},
  \bibinfo{author}{\bibfnamefont{J.~W.} \bibnamefont{Lynn}},
  \bibinfo{author}{\bibfnamefont{J.}~\bibnamefont{Hu}},
  \bibinfo{author}{\bibfnamefont{Y.-L.} \bibnamefont{Huang}},
  \bibinfo{author}{\bibfnamefont{F.-C.} \bibnamefont{Hsu}},
  \bibinfo{author}{\bibfnamefont{K.-W.} \bibnamefont{Yeh}},
  \bibinfo{author}{\bibfnamefont{M.}~\bibnamefont{kuen Wu}},
  \bibnamefont{et~al.}, \bibinfo{journal}{Phys. Rev. B}
  \textbf{\bibinfo{volume}{79}}, \bibinfo{pages}{054503}
  (\bibinfo{year}{2009}).

\bibitem[{\citenamefont{Bao et~al.}(2009)\citenamefont{Bao, Qiu, Huang, Green,
  Zajdel, Fitzsimmons, Zhernenkov, Chang, Fang, Qian et~al.}}]{bao-2008}
\bibinfo{author}{\bibfnamefont{W.}~\bibnamefont{Bao}},
  \bibinfo{author}{\bibfnamefont{Y.}~\bibnamefont{Qiu}},
  \bibinfo{author}{\bibfnamefont{Q.}~\bibnamefont{Huang}},
  \bibinfo{author}{\bibfnamefont{M.~A.} \bibnamefont{Green}},
  \bibinfo{author}{\bibfnamefont{P.}~\bibnamefont{Zajdel}},
  \bibinfo{author}{\bibfnamefont{M.~R.} \bibnamefont{Fitzsimmons}},
  \bibinfo{author}{\bibfnamefont{M.}~\bibnamefont{Zhernenkov}},
  \bibinfo{author}{\bibfnamefont{S.}~\bibnamefont{Chang}},
  \bibinfo{author}{\bibfnamefont{M.}~\bibnamefont{Fang}},
  \bibinfo{author}{\bibfnamefont{B.}~\bibnamefont{Qian}}, \bibnamefont{et~al.},
  \bibinfo{journal}{Phys. Rev. Lett.} \textbf{\bibinfo{volume}{102}},
  \bibinfo{pages}{247001} (\bibinfo{year}{2009}).

\bibitem[{\citenamefont{Zhao et~al.}(2008{\natexlab{b}})\citenamefont{Zhao, Ratcliff,
  Lynn, Chen, Luo, Wang, Hu, and Dai}}]{zhao:140504}
\bibinfo{author}{\bibfnamefont{J.}~\bibnamefont{Zhao}},
  \bibinfo{author}{\bibfnamefont{W.} \bibnamefont{Ratcliff}},
  \bibinfo{author}{\bibfnamefont{J.~W.} \bibnamefont{Lynn}},
  \bibinfo{author}{\bibfnamefont{G.~F.} \bibnamefont{Chen}},
  \bibinfo{author}{\bibfnamefont{J.~L.} \bibnamefont{Luo}},
  \bibinfo{author}{\bibfnamefont{N.~L.} \bibnamefont{Wang}},
  \bibinfo{author}{\bibfnamefont{J.}~\bibnamefont{Hu}}, \bibnamefont{and}
  \bibinfo{author}{\bibfnamefont{P.}~\bibnamefont{Dai}},
  \bibinfo{journal}{Phys. Rev. B} \textbf{\bibinfo{volume}{78}},
  \bibinfo{pages}{140504(R)} (\bibinfo{year}{2008}{\natexlab{b}}).

\bibitem[{\citenamefont{Maier and Scalapino}(2008)}]{maier:020514}
\bibinfo{author}{\bibfnamefont{T.~A.} \bibnamefont{Maier}} \bibnamefont{and}
  \bibinfo{author}{\bibfnamefont{D.~J.} \bibnamefont{Scalapino}},
  \bibinfo{journal}{Phys. Rev. B} \textbf{\bibinfo{volume}{78}},
  \bibinfo{pages}{020514(R)} (\bibinfo{year}{2008}).

\bibitem[{\citenamefont{Yin et~al.}(2008)\citenamefont{Yin, Leb\`{e}gue, Han,
  Neal, Savrasov, and Pickett}}]{yin:047001}
\bibinfo{author}{\bibfnamefont{Z.~P.} \bibnamefont{Yin}},
  \bibinfo{author}{\bibfnamefont{S.}~\bibnamefont{Leb\`{e}gue}},
  \bibinfo{author}{\bibfnamefont{M.~J.} \bibnamefont{Han}},
  \bibinfo{author}{\bibfnamefont{B.~P.} \bibnamefont{Neal}},
  \bibinfo{author}{\bibfnamefont{S.~Y.} \bibnamefont{Savrasov}},
  \bibnamefont{and} \bibinfo{author}{\bibfnamefont{W.~E.}
  \bibnamefont{Pickett}}, \bibinfo{journal}{Phys. Rev. Lett.}
  \textbf{\bibinfo{volume}{101}}, \bibinfo{pages}{047001}
  (\bibinfo{year}{2008}).

\bibitem[{\citenamefont{Ma et~al.}(2009)\citenamefont{Ma, Ji, Hu, Lu, and
  Xiang}}]{ma-2009}
\bibinfo{author}{\bibfnamefont{F.}~\bibnamefont{Ma}},
  \bibinfo{author}{\bibfnamefont{W.}~\bibnamefont{Ji}},
  \bibinfo{author}{\bibfnamefont{J.}~\bibnamefont{Hu}},
  \bibinfo{author}{\bibfnamefont{Z.-Y.} \bibnamefont{Lu}}, \bibnamefont{and}
  \bibinfo{author}{\bibfnamefont{T.}~\bibnamefont{Xiang}},
  \bibinfo{journal}{Phys. Rev. Lett.} \textbf{\bibinfo{volume}{102}},
  \bibinfo{eid}{177003} (\bibinfo{year}{2009}).

\bibitem[{\citenamefont{Johannes and Mazin}(2009)}]{johannes-2009}
\bibinfo{author}{\bibfnamefont{M.~D.} \bibnamefont{Johannes}} \bibnamefont{and}
  \bibinfo{author}{\bibfnamefont{I.~I.} \bibnamefont{Mazin}},
  \bibinfo{journal}{Phys. Rev. B} \textbf{\bibinfo{volume}{79}},
  \bibinfo{eid}{220510(R)} (\bibinfo{year}{2009}).

\bibitem[{\citenamefont{Fang et~al.}(2009{\natexlab{b}})\citenamefont{Fang,
  Bernevig, and Hu}}]{fang-2009}
\bibinfo{author}{\bibfnamefont{C.}~\bibnamefont{Fang}},
  \bibinfo{author}{\bibfnamefont{B.~A.} \bibnamefont{Bernevig}},
  \bibnamefont{and} \bibinfo{author}{\bibfnamefont{J.}~\bibnamefont{Hu}},
  \bibinfo{journal}{Europhys. Lett.} \textbf{\bibinfo{volume}{86}},
  \bibinfo{pages}{67005} (\bibinfo{year}{2009}{\natexlab{b}}).

\bibitem[{\citenamefont{Guinier}(1994)}]{guinier-1994}
\bibinfo{author}{\bibfnamefont{A.}~\bibnamefont{Guinier}},
  \emph{\bibinfo{title}{X-Ray Diffraction in Crystals, Imperfect Crystals, and
  Amorphous Bodies}} (\bibinfo{publisher}{Dover}, \bibinfo{address}{New York},
  \bibinfo{year}{1994}), Chap.~\bibinfo{chapter}{9}.

\bibitem[{\citenamefont{Qiu et~al.}()\citenamefont{Qiu, Bao, Zhao, Broholm,
  Stanev, Tesanovic, Gasparovic, Chang, Hu, Qian et~al.}}]{qiu-2009}
\bibinfo{author}{\bibfnamefont{Y.}~\bibnamefont{Qiu}},
  \bibinfo{author}{\bibfnamefont{W.}~\bibnamefont{Bao}},
  \bibinfo{author}{\bibfnamefont{Y.}~\bibnamefont{Zhao}},
  \bibinfo{author}{\bibfnamefont{C.}~\bibnamefont{Broholm}},
  \bibinfo{author}{\bibfnamefont{V.}~\bibnamefont{Stanev}},
  \bibinfo{author}{\bibfnamefont{Z.}~\bibnamefont{Tesanovic}},
  \bibinfo{author}{\bibfnamefont{Y.~C.} \bibnamefont{Gasparovic}},
  \bibinfo{author}{\bibfnamefont{S.}~\bibnamefont{Chang}},
  \bibinfo{author}{\bibfnamefont{J.}~\bibnamefont{Hu}},
  \bibinfo{author}{\bibfnamefont{B.}~\bibnamefont{Qian}}, \bibnamefont{et~al.},
  \bibinfo{journal}{Phys. Rev. Lett.} \textbf{\bibinfo{volume}{103}},
  \bibinfo{pages}{067008} (\bibinfo{year}{2009}{\natexlab{b}}).

\bibitem[{\citenamefont{Wen et~al.}()\citenamefont{Wen, Xu, Xu, Chen, Chi, Gu, and
  Tranquada}}]{wen-2009}
\bibinfo{author}{\bibfnamefont{Jinsheng}~\bibnamefont{Wen}},
  \bibinfo{author}{\bibfnamefont{Guangyong}~\bibnamefont{Xu}},
  \bibinfo{author}{\bibfnamefont{Zhijun}~\bibnamefont{Xu}},
  \bibinfo{author}{\bibfnamefont{Ying}~\bibnamefont{Chen}},
  \bibinfo{author}{\bibfnamefont{Songxue}~\bibnamefont{Chi}},
  \bibinfo{author}{\bibfnamefont{Genda}~\bibnamefont{Gu}}, \bibnamefont{and}
  \bibinfo{author}{\bibfnamefont{J.~M.} \bibnamefont{Tranquada}},
  \bibinfo{note}{(unpublished)}.

\bibitem[{\citenamefont{Demler et~al.}(2001)\citenamefont{Demler, Sachdev, and
  Zhang}}]{prl67202}
\bibinfo{author}{\bibfnamefont{E.}~\bibnamefont{Demler}},
  \bibinfo{author}{\bibfnamefont{S.}~\bibnamefont{Sachdev}}, \bibnamefont{and}
  \bibinfo{author}{\bibfnamefont{Y.}~\bibnamefont{Zhang}},
  \bibinfo{journal}{Phys. Rev. Lett.} \textbf{\bibinfo{volume}{87}},
  \bibinfo{pages}{067202} (\bibinfo{year}{2001}).

\bibitem[{\citenamefont{Moodenbaugh et~al.}(1988)\citenamefont{Moodenbaugh, Xu,
  Suenaga, Folkerts, and Shelton}}]{moodenbaugh}
\bibinfo{author}{\bibfnamefont{A.~R.} \bibnamefont{Moodenbaugh}},
  \bibinfo{author}{\bibfnamefont{Y.}~\bibnamefont{Xu}},
  \bibinfo{author}{\bibfnamefont{M.}~\bibnamefont{Suenaga}},
  \bibinfo{author}{\bibfnamefont{T.~J.} \bibnamefont{Folkerts}},
  \bibnamefont{and} \bibinfo{author}{\bibfnamefont{R.~N.}
  \bibnamefont{Shelton}}, \bibinfo{journal}{Phys. Rev. B}
  \textbf{\bibinfo{volume}{38}}, \bibinfo{pages}{4596} (\bibinfo{year}{1988}).

\bibitem[{\citenamefont{Lake et~al.}(2005)\citenamefont{Lake, Lefmann,
  Christensen, Aeppli, Mcmorrow, R\o{}nnow, Vorderwisch, Smeibidl,
  Mangkorntong, Sasagawa et~al.}}]{lakefield}
\bibinfo{author}{\bibfnamefont{B.}~\bibnamefont{Lake}},
  \bibinfo{author}{\bibfnamefont{K.}~\bibnamefont{Lefmann}},
  \bibinfo{author}{\bibfnamefont{N.~B.} \bibnamefont{Christensen}},
  \bibinfo{author}{\bibfnamefont{G.}~\bibnamefont{Aeppli}},
  \bibinfo{author}{\bibfnamefont{D.~F.} \bibnamefont{Mcmorrow}},
  \bibinfo{author}{\bibfnamefont{H.~M.} \bibnamefont{R\o{}nnow}},
  \bibinfo{author}{\bibfnamefont{P.}~\bibnamefont{Vorderwisch}},
  \bibinfo{author}{\bibfnamefont{P.}~\bibnamefont{Smeibidl}},
  \bibinfo{author}{\bibfnamefont{N.}~\bibnamefont{Mangkorntong}},
  \bibinfo{author}{\bibfnamefont{T.}~\bibnamefont{Sasagawa}},
  \bibnamefont{et~al.}, \bibinfo{journal}{Nature Mater.}
  \textbf{\bibinfo{volume}{4}}, \bibinfo{pages}{658} (\bibinfo{year}{2005}).

\bibitem[{\citenamefont{Zhang et~al.}(2009)\citenamefont{Zhang, Singh, and
  Du}}]{zhang:012506}
\bibinfo{author}{\bibfnamefont{L.}~\bibnamefont{Zhang}},
  \bibinfo{author}{\bibfnamefont{D.~J.} \bibnamefont{Singh}}, \bibnamefont{and}
  \bibinfo{author}{\bibfnamefont{M.~H.} \bibnamefont{Du}},
  \bibinfo{journal}{Phys. Rev. B} \textbf{\bibinfo{volume}{79}},
  \bibinfo{pages}{012506} (\bibinfo{year}{2009}).

\bibitem[{\citenamefont{McQueen et~al.}(2009)\citenamefont{McQueen, Huang,
  Ksenofontov, Felser, Xu, Zandbergen, Hor, Allred, Williams, Qu
  et~al.}}]{mcqueen:014522}
\bibinfo{author}{\bibfnamefont{T.~M.} \bibnamefont{McQueen}},
  \bibinfo{author}{\bibfnamefont{Q.}~\bibnamefont{Huang}},
  \bibinfo{author}{\bibfnamefont{V.}~\bibnamefont{Ksenofontov}},
  \bibinfo{author}{\bibfnamefont{C.}~\bibnamefont{Felser}},
  \bibinfo{author}{\bibfnamefont{Q.}~\bibnamefont{Xu}},
  \bibinfo{author}{\bibfnamefont{H.}~\bibnamefont{Zandbergen}},
  \bibinfo{author}{\bibfnamefont{Y.~S.} \bibnamefont{Hor}},
  \bibinfo{author}{\bibfnamefont{J.}~\bibnamefont{Allred}},
  \bibinfo{author}{\bibfnamefont{A.~J.} \bibnamefont{Williams}},
  \bibinfo{author}{\bibfnamefont{D.}~\bibnamefont{Qu}}, \bibnamefont{et~al.},
  \bibinfo{journal}{Phys. Rev. B} \textbf{\bibinfo{volume}{79}},
  \bibinfo{pages}{014522} (\bibinfo{year}{2009}).

\bibitem[{\citenamefont{Liu et~al.}()\citenamefont{Liu, Ke, Qian, Hu, Fobes,
  Vehstedt, Pham, Yang, Fang, Spinu et~al.}}]{liu}
\bibinfo{author}{\bibfnamefont{T.~J.} \bibnamefont{Liu}},
  \bibinfo{author}{\bibfnamefont{X.}~\bibnamefont{Ke}},
  \bibinfo{author}{\bibfnamefont{B.}~\bibnamefont{Qian}},
  \bibinfo{author}{\bibfnamefont{J.}~\bibnamefont{Hu}},
  \bibinfo{author}{\bibfnamefont{D.}~\bibnamefont{Fobes}},
  \bibinfo{author}{\bibfnamefont{E.~K.} \bibnamefont{Vehstedt}},
  \bibinfo{author}{\bibfnamefont{H.}~\bibnamefont{Pham}},
  \bibinfo{author}{\bibfnamefont{J.~H.} \bibnamefont{Yang}},
  \bibinfo{author}{\bibfnamefont{M.~H.} \bibnamefont{Fang}},
  \bibinfo{author}{\bibfnamefont{L.}~\bibnamefont{Spinu}},
  \bibnamefont{et~al.},
  \eprint{arXiv:0904.0824}.

\bibitem[{\citenamefont{Mook et~al.}()\citenamefont{Mook, Lumsden,
  Christianson, Sales, Jin, McGuire, Sefat, Mandrus, Nagler, Egami
  et~al.}}]{mook-2009}
\bibinfo{author}{\bibfnamefont{H.~A.} \bibnamefont{Mook}},
  \bibinfo{author}{\bibfnamefont{M.}~\bibnamefont{Lumsden}},
  \bibinfo{author}{\bibfnamefont{A.}~\bibnamefont{Christianson}},
  \bibinfo{author}{\bibfnamefont{B.~C.} \bibnamefont{Sales}},
  \bibinfo{author}{\bibfnamefont{R.}~\bibnamefont{Jin}},
  \bibinfo{author}{\bibfnamefont{M.~A.} \bibnamefont{McGuire}},
  \bibinfo{author}{\bibfnamefont{A.}~\bibnamefont{Sefat}},
  \bibinfo{author}{\bibfnamefont{D.}~\bibnamefont{Mandrus}},
  \bibinfo{author}{\bibfnamefont{S.}~\bibnamefont{Nagler}},
  \bibinfo{author}{\bibfnamefont{T.}~\bibnamefont{Egami}},
  \bibnamefont{et~al.}, \eprint{arXiv:0904.2178}.

\end{thebibliography}

\end{document}